\documentclass[traditabstract]{aa}
\usepackage{txfonts} 
\usepackage{natbib}
\bibpunct{(}{)}{;}{a}{}{,} 
\usepackage{graphicx}

\begin{document}

\title{Differentially-rotating neutron star models with a parametrized
rotation profile} \titlerunning{Differentially rotating neutron stars}

\author{Filippo Galeazzi\inst{1}\thanks{filippo.galeazzi@aei.mpg.de},
  Shin'ichirou Yoshida\inst{2},  Yoshiharu Eriguchi\inst{2} }

\institute{Max-Planck-Institut f\"{u}r Gravitationsphysik,
  Albert-Einstein-Institut, Am M\"{u}hlenberg 1, D-14476 Potsdam,
  Deutschland \and Department of Earth Science and Astronomy, Graduate
  School of Arts and Sciences, University of Tokyo, Komaba, Meguro-ku
  3-8-1, 153-8902 Tokyo, Japan}

\date{Recieved / Accepted}

\abstract {We analyze the impact of the choice rotation law on
  equilibrium sequences of relativistic differentially-rotating
  neutron stars in axisymmetry. The maximum allowed mass for each
  model is strongly affected by the distribution of angular
  velocity along the radial direction and by the consequent degree of
  differential rotation. In order to study the wide parameter space
  implied by the choice of rotation law, we introduce a functional
  form that generalizes the so called ``j-const. law" adopted in all
  previous work. 
  Using this new rotation law we reproduce the angular velocity 
  profile of differentially-rotating remnants from the coalescence of 
  binary neutron stars in various 3-dimensional 
  dynamical simulations.
  We compute equilibrium sequences of differentially rotating stars 
  with a polytropic equation of state starting
  from the spherically symmetric static case. By analyzing the sequences 
 at constant ratio, $T/|W|$, of rotational kinetic energy to gravitational 
 binding energy, we find that the parameters that best describe the 
 binary neutron star remnants cannot produce equilibrium configurations
 with values of $T/|W|$ that exceed $0.14$, the criterion for  the onset of the 
 \emph{secular} instability. 
  }

\maketitle
\section{Introduction}
A neutron star (NS), during most of its life, is considered to be a
stationary and rigidly rotating object, apart from a tiny lag between 
the rotation of the superfluid component and that of the normal fluid and 
the crust (e.g. \citet{Baym1969} and \citet{Pines1985}).

In fact a nascent neutron star which is born in a supernova event 
is likely to rotate differentially at first before its angular velocity distribution 
evolves toward a uniform rotation. 
There are several mechanisms that
account for the redistribution of the angular momentum.
One such mechanism is the shear viscosity of neutron matter \citep{Sawyer1989}.  
An estimate of the timescale in
which the viscosity damps a neutron star's internal shear motion has
received a lot of attention in the literature (e.g. \citet{Cutler1987}).
 For a typical neutron star we expect a timescale of $10-100$ yrs.
 In the presence of magnetic fields, the magnetic braking \citep{Spruit1999} or the
magnetorotational instability (MRI) \citep{Balbus1991} may 
drastically accelerate the redistribution of angular momentum
to the order of 1s \citep{Duez2006}.

Differential rotation plays an important role
in the beginning and at the end of the life of a NS.
 In ots early life, strong differential rotation of a massive core
in a supernova may affect the collapse and bounce 
dynamics \citep{Dimmelmeier2002, Ott2004}.
A newly-born neutron star with strong differential
rotation may lose its stability in the course of the rotational (and thermal)
evolution, and may collapse to a black hole or other type of compact
star (quark star, hybrid star). This may give a unique neutrino signal 
and a strong gravitational wave emission from the supernova explosion 
of massive stars (see, e.g., \citet{Ott2007} and \citet{Kotake2010}
for recent studies on gravitational wave mechanism from stellar core collapse).

An exciting way in which a neutron star's life can end is in 
the inspiral and consequent merger with a binary companion.
Recent results of numerical
3-dimensional simulations of neutron star binary mergers
(e.g., \citet{Ruffert1996};  \citet{Ruffert2001}; 
\citet{ShibataUryu2002}; \citet{Shibata2006};  \citet{Anderson2008};
\citet{Baiotti2008}; \citet{Giacomazzo2010}) have shown
that there are cases in which 
the remnant of the merger has a significantly high degree of
differential rotation such that it can sustain a total mass
considerably larger than that of a uniformly rotating 
star \citep{Baiotti2008}. These hyper-massive neutron stars
(HMNSs) remain stable over many dynamical
timescales before collapsing to a BH (e.g. \citet{ShibataUryu2002},
\citet{Baiotti2008}).  
In order to study
the effect of the centrifugal force in supporting the HMNS, most of these
papers show the angular velocity profile of the merged object before the
eventual collapse to a black hole.  The typical profile extracted from
simulations characteristically shows a plateau near the rotation
axis and a certain distance from the axis shows a nearly power law
behavior. The power index seems to differ from simulation to simulation,
but  generally does not agree with the so-called ``j-const. law" \citep{EriguchiMueller1985}
of rotation, which has been so far the only choice available to
construct equilibrium models of differentially rotating stars.

Models of rapidly rotating stars in general relativity have been studied
since the 70s, when large numerical computing facilities became
available. The centrifugal deformation and general relativistic gravity
make these investigations fully reliant on numerical methods. 
Since the pioneering work of \citet{Ipser1976},
these studies have included progressively more sophisticated aspects, 
such as nuclear matter EOS, degrees of differential rotation, magnetic fields
and quite recently meridional flows \citep{Birkl_etal2010}.
We here
name the following references of such studies:
 \citet{Ipser1976} ; \citet{Komatsu1989a} 
  ; \citet{Bonazzola1993} ; 
\citet{StergioulasFriedman1995} ; \citet{Shapiro2000} ; \citet{Ansorg_etal2002}.
For more comprehensive collection of literatures, 
see \citet{Stergioulas_LivingReview}.

However, it is surprising to note that all of these studies 
have used just a single kind of ansatz on the rotational angular velocity
profile (namely the ``j-const. law", which include uniformly-rotating stars
as a limiting case (Section \ref{subsec:rotation profile})). 

In this work we introduce a new parametrized functional form of the 
rotation profile that enables us to investigate a broader class
of differentially rotating stars. In particular, our new rotation law reproduces
different power laws in the outer envelope of the angular velocity distribution (see Section\ref{subsec:rotation profile}) with which it is possible to match the profile of the HMNS . 
To analyze the impact  of the law of rotation on the 
maximum allowed mass for the differentially-rotating neutron star we construct equilibrium  sequences by imposing a fixed value of $T/|W|$ close to what is the classical limit for the onset of the \emph{secular} instability; rotating neutron stars are known to be destabilized by the effect of
dissipative mechanisms like viscosity and gravitational radiation reaction 
(Chandrasekhar-Friedman-Schutz (CFS) instability; 
\citet{Chandrasekhar1970,FriedmanSchutz1978}). 
We show that the classical secular instability criterion is satisfied only for a
limited class of the rotation profiles.

The paper is organized as follows. In Section \ref{sec::formulation} we briefly review the
formalism of the equations and the numerical method used to solve
them. Then we describe the new functional form of the rotation profile.
In Section \ref{sec::results} we construct 
equilibrium sequences of rotating stars using the new rotation profile.
Discussions are given in the final section.
\section{Formulation}\label{sec::formulation}
\subsection{Equations for fluid and spacetime}
We construct configurations for an \emph{axisymmetric} and \emph{stationary} 
rotating perfect fluid in general relativity.
The spacetime is assumed to be asymptotically flat and 
the flow is assumed to be circular (the velocity is only in the azimuthal direction).
In this case we have two Killing vectors \citep{Bardeen1970}
and can choose a coordinate system in such a way that the metric of the spacetime
is written as (e.g., \citet{Komatsu1989a}),
\begin{equation}
  ds^2 = -e^{2\nu}dt^2 
  + e^{2\alpha} \left(dr^2+r^2d\theta^2\right)
  + e^{2\beta}r^2\sin^2\theta(d\varphi-\omega dt)^2,
\end{equation}
where the spherical polar coordinates $(r,\theta,\phi)$ are used.
The metric potentials $\alpha,\beta,\nu$ and $\omega$ are
functions of $r$ and $\theta$ only.
\footnote{We use geometrized units, $c=G=1$.}
The energy momentum tensor $T^{ab}$ of a perfect fluid is 
\begin{equation}
T^{ab}=(\epsilon + p) u^a u ^b+ p g^{ab}
\end{equation}
where $\epsilon$, $p$ and $u^a$ are the total energy density, the pressure 
and the four velocity, respectively. 

The basic equations are: 1) Einstein's equation for the metric potentials
$G_{ab}=8\pi T_{ab}$, 2) rest mass conservation $\nabla_a(\rho u^a)=0$,
which is trivially satisfied under the present assumptions 
and 3) stress-energy conservation $\nabla_bT^{ab}=0$.

As described in \citet{Komatsu1989a}, the components of Einstein's equation 
are cast into 4 equations for the potentials,
three of which are elliptic partial differential equations. They are transformed
into convenient integral equations by using appropriate Green's functions.

The spatial linear velocity $V$ of the flow with respect to an observer with zero angular momentum is given
by
\begin{equation}
	V = e^{\beta-\nu}r\sin\theta (\Omega-\omega)
\end{equation}
where $\Omega=u^\varphi/u^t$ is the angular frequency of the fluid measured in the
asymptotic inertial frame. We next introduce the specific angular momentum
\footnote{This quantity is one definition of specific angular momentum
in general relativity. For different definitions, see \citet{Kozlowski_etal1978}.} $j$:
\begin{equation}
	j = -\frac{u_\varphi}{u_t}.
\end{equation}
Using these quantities, we can write the equilibrium equations for a stationary configuration 
as
\begin{equation}
	\frac{\partial_A p}{\epsilon +p} + \partial_A\nu
	- \frac{1}{1-V^2}V\partial_A V + \frac{j}{1-j\Omega}\partial_A\Omega = 0
	\quad (A=r,\theta).
	\label{hydrostatic eq}
\end{equation}
Assuming the fluid to be barotropic, the condition of integrability for Eq.(\ref{hydrostatic eq}),
can be written as
\begin{equation}
	\frac{j}{1-j\Omega} = g(\Omega),
	\label{integrability of hydrostat}
\end{equation}
by introducing an arbitrary functional $g$, which is detailed in the next subsection.
The first integral of motion for a stationary solution can be written as 
\begin{equation}
	\int\frac{dp}{\epsilon +p} + \nu +
	\frac{1}{2}\ln(1-V^2)+ \int g(\Omega)d\Omega = 0.
	\label{1st integral of hydrostatic eq}
\end{equation}

The components of Einstein's equation and the first integral of hydrostatic equation
can be iteratively solved as described in \citet{Komatsu1989a}

\subsection{Rotation profile\label{subsec:rotation profile}}
To exploit the first integral of the hydrostatic equation, we need to
impose the integrability condition for Eq.(\ref{integrability of hydrostat}).
Different choices of functional form $g(\Omega)$
would lead to various classes of differential-rotation profiles.
All previous studies of differentially-rotating relativistic stellar
models assume 
the simplest linear functional form,

\begin{equation}
  \label{eq:jlaw}
  g(\Omega)=A^2(\Omega_c-\Omega),
\end{equation}
where $A$ is a constant with the dimensions of a length and $\Omega_c$ is the angular velocity on the axis of rotation.

This choice is termed as ``j-const. law" since the Newtonian limit of
the specific angular momentum $j$ is that of ``j-const. law" in
Newtonian rotating stars \citep{EriguchiMueller1985}.

We here introduce a more flexible form of the rotation profile as
\begin{equation}
	g(\Omega) = \frac{\frac{R_0^2}{\Omega_c^\alpha}\Omega
	(\Omega^\alpha-\Omega_c^\alpha)}
	{1-\frac{R_0^2}{\Omega_c^\alpha}\Omega^2
	(\Omega^\alpha-\Omega_c^\alpha)},
	\label{eq: functional form of g}
\end{equation}
where $\alpha$, $R_0$ and $\Omega_c$ are constants.
The corresponding specific angular momentum is
\begin{equation}
	j = \frac{R_0^2}{\Omega_c^\alpha}\Omega 
	(\Omega^\alpha-\Omega_c^\alpha).
\end{equation}

The physical significance of the parameters $\alpha$, $R_0$ and $\Omega_c$ in the profile is easily seen
if we consider its Newtonian limit. Then the angular frequency is written as
\begin{equation}
	\Omega = \Omega_c
	\left(1+\left(\frac{R}{R_0}\right)^2\right)^{\frac{1}{\alpha}}
\end{equation}
where $R=r\sin\theta$. For $R \ll R_0$, we have $\Omega\sim\Omega_c$,
while for $R\gg R_0$,
\begin{equation}
	\Omega \sim \Omega_c\left(\frac{R}{R_0}\right)^{\frac{2}{\alpha}}.
\end{equation}
This means that the rotation profile has an inner plateau inside $R\sim R_0$
and a power law envelope for $R\gg R_0$. The introduction of the index $\alpha$ is 
an important feature of the new profile: as shown in
Table 1, it is possible to reproduce different rotation
laws by choosing different values for $\alpha$.


\begin{table}%
\centering
\vspace{2 mm}
  \begin{center}
    \begin{tabular}{c||c|c|c|c}\hline
      $\alpha$ & -1 & -4/3 & -2 & -4\\
      $\Omega_{outer}$ & $R^{-2}$ & $R^{-3/2}$ & $R^{-1}$ & $R^{-1/2}$\\
           power law in envelope & j-const. & Keplerian & v-const. & HMNS \\
      \hline
    \end{tabular}
    \vspace{1 mm}
    \caption{Choosing different $\alpha$, it is possible to introduce different power
    law distributions of the angular velocity in the outer region of a star. We here
    tabulate the Newtonian limits of particular choices of $\alpha$.
   ``j-const." refers to constant specific angular momentum.``v-const." is for
    constant linear velocity. ``HMNS" is the approximate power law of  some 
    of numerical relativistic simulations with HMNS formation (e.g., \citet{ShibataUryu2002}; \citet{Baiotti2008}).}
  \end{center}
    \label{tab:laws}
\end{table}
For some interesting cases ($\alpha \in \left\{-1,-2,-4\right\} $), it
is possible to integrate analytically the expression for $g(\Omega)$
(see Appendix).
It is important to note that for $\alpha=-1$ the $j$-constant type of
law is recovered  -- the two functional
forms must agree in the limit of weak gravity. This particular case is marginally stable  under Rayleigh's local stability criterion of axisymmetric instability, which states that the specific angular momentum must not decrease outward in a stable star.
Values of $\alpha$ smaller than $-1$ satisfy this condition (and hence are stable). $\alpha > 0$
also satisfies this condition, but a star cannot be spun up rapidly enough
to allow the appearance of the CFS instability, because it easily reaches its mass-shedding
limit.

We now apply the new rotation law to construct equilibrium
configurations of differentially rotating neutron stars and study
their dependence on $\alpha$. Each configuration is defined by five
parameters: the axis ratio $r_p/r_e$, the polytropic index $N$, the
maximum density $ \rho_{max}$ and the two parameters of the rotation law,
$R_0$ and $\alpha$.

\section{Results}\label{sec::results}
\subsection{Numerical code}
We construct sequences of axisymmetric and differentially rotating
objects using the code discussed in \citet{Komatsu1989a}. The code
iteratively solves Einstein's equation and the first integral of motion for
a stationary fluid configuration.

For all computation we used a grid with 600 points in the $r-$ direction
and 300 in the $\theta-$ direction. The radial grid is set up in such a way that 
the inner uniformly-distributed 300 grid points just cover the whole star (on the equator the 
300th point from the origin always corresponds to the surface of
the star).  Outside the star we adopt the ``compactified" coordinate
as in \citet{Cook1992}, to take into account the exact boundary condition
at the radial infinity.

For a given EOS, a functional of the rotation profile and maximum density,
the code requires information on "how rapidly" the star is rotating. To this end, it is more
convenient to specify a parameter that measures the deformation induced
by the rotation rather than the angular momentum or the rotational frequency.
We follow \citet{Komatsu1989a} in fixing the ratio of the polar coordinate radius of the star
to the equatorial radius. The physical parameters of rotation as well as other
physical quantities are then computed.

\subsection{Equation of state}
In the current study we use a polytropic equation of state, with the index $N=1$ to model neutron stars, 
\begin{equation}
p=\kappa \rho^{1+1/N}, \epsilon=\rho+Np.
\end{equation}
Typical neutron star masses and radi are recovered
when we set $\kappa=100$ in geometrized units. We fix this parameter
for all the sequences to have a crude model of the nuclear equation of state
for neutron matter. In a subsequent paper we will investigate the dependence of
the space of configurations on the EOS, using realistic cold and 
finite-temperature EOSs.

\subsection{Parameters}
To obtain an equilibrium model, we fix the parameters of the rotation profile,
$\alpha$ and $R_0$, then choose the maximum density $\rho_{\rm  max}$
and the coordinate axis ratio. In our code this set of parameters uniquely 
determines an equilibrium configuration, whose physical characteristics such
as mass, compactness and angular momentum, angular frequency at the centre
 are computed once the solution is obtained. Rest mass and gravitational mass 
are computed by using the standard formula (see e.g. \citet{Bardeen1970}). 
An important measure of rotation, the $T/|W|$
parameter, is introduced as in \citet{Komatsu1989a}. This is used both
in Newtonian and general relativistic studies of rotating stars and defined
as the ratio of rotational kinetic energy to gravitational binding energy
(see \citet{Komatsu1989a} for details). It characterises the overall
strength of rotation of the star. Classical studies suggest that when
$T/|W|\sim 0.14$, bar-shaped equilibria bifurcate 
(into "Jacobi" or "Dedekind" type equilibrium sequences)  and
$T/|W|\sim 0.27$ marks the onset of dynamical instability of axisymmetric
configurations \citep{Tassoul1978}. In general relativity numerical simulations 
point to a rather lower limit, $T/|W|\sim 0.25$ \citep{Manca2007}.

\subsection{Numerical equilibrium sequences
\label{subsec: numerical sequence}}
For a given rotation profile and a maximum density, we start from a non-rotating 
model (the polar-to-equatorial axis ratio is equal to unity), and decrease the
ratio to obtain stars with increasingly rapid rotation.  We terminate the sequence
if one of  the following conditions is satisfied:
1) $T/|W| = 0.14$. This is the classical criterion at which CFS
instability or viscous instability sets in for a bar-shaped deformation
(the actual critical point depends weakly on both the equation of state and degree
of differential rotation \citep{Karino2002}).
2) Mass-shedding limit. The last condition is reached when the angular frequency
 of matter at the equatorial surface comes close to local Keplerian frequency and a 
further spin-up of a star leads to shedding of mass. Beyond this point it is not possible to 
construct equilibrium configurations.
3) Topology change. When the degree of differential rotation is large the maximum density
may move away from the rotation axis. When the
value of the axis ratio reaches zero, the star develops a hole on the rotation
axis. The equilibrium sequence may be continued beyond this point, but
we are not interested in this toroidal shape configuration as a model of 
a neutron star. 

Concerning the last point above, we should note that
the structure of the parameter space is quite complex, as has been
analyzed by \citet{Ansorg2009}. The toroidal and the spheroidal
configurations form disjoint families of solutions separated by a mass
shedding region. Increasing the degree of differential rotation by
reducing the value of the $R_0$ parameter, the two families join again 
in the parameter space.  The area of the region separating the two topologically 
different families depends on the $\rho_{max}$ of the configuration, becoming smaller
as the latter is increasing. Stated differently, starting from the Newtonian limit and 
going to the more general relativistic case, it becomes more and more complicated 
to find a value of $R_0$ for which the solution can reach the toroidal family, in fact 
the separation increases between the two types of solutions. We have 
studied the parameter space of the solutions
for the three different rotation laws analyzed here, and will collect the
results along with the dependency hidden in the EOS in the followup paper.

\subsection{Stellar mass in parameter space}

\begin{figure}[hbtp]
\begin{center}
  \includegraphics[scale=0.45]{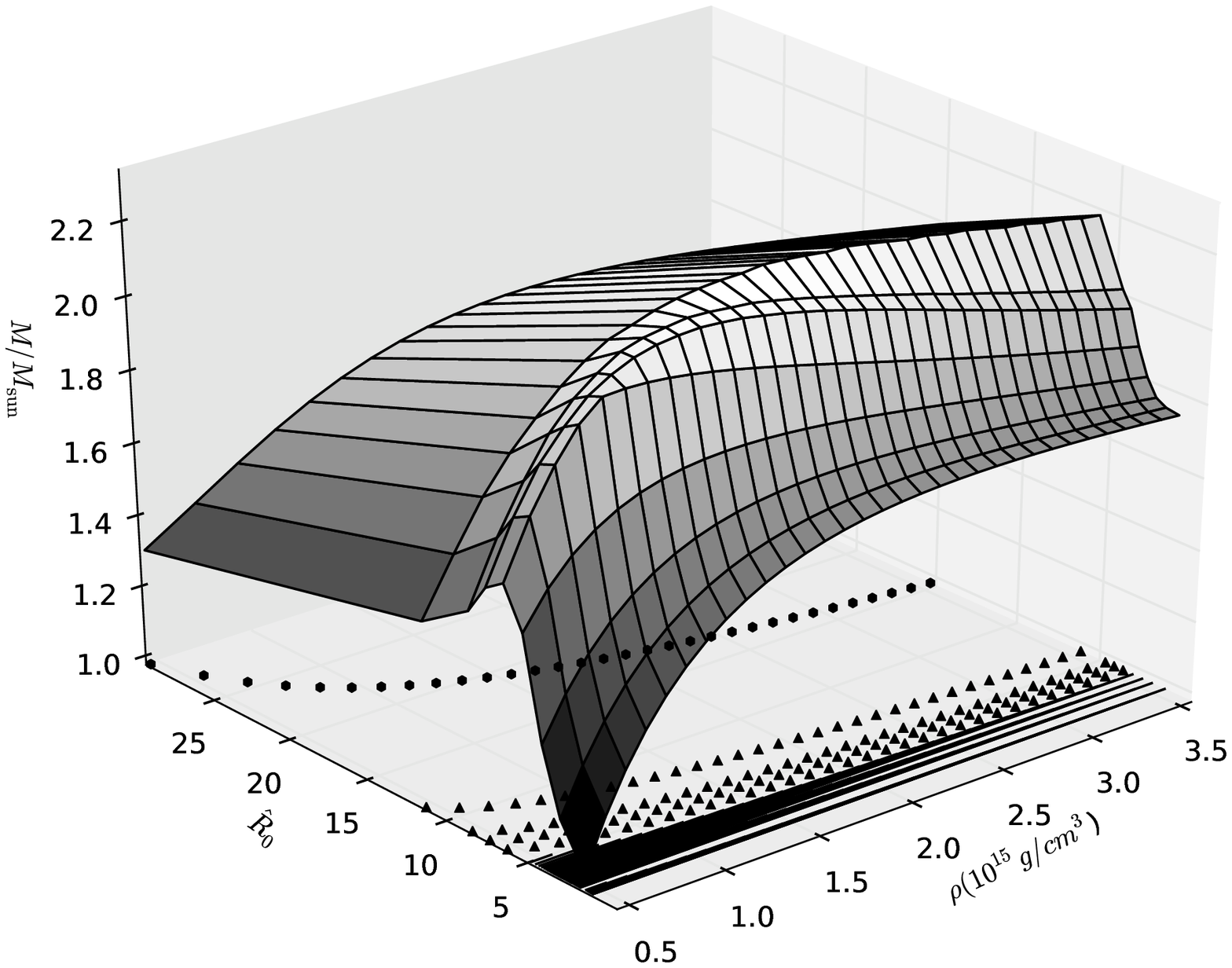}
  \includegraphics[scale=0.45]{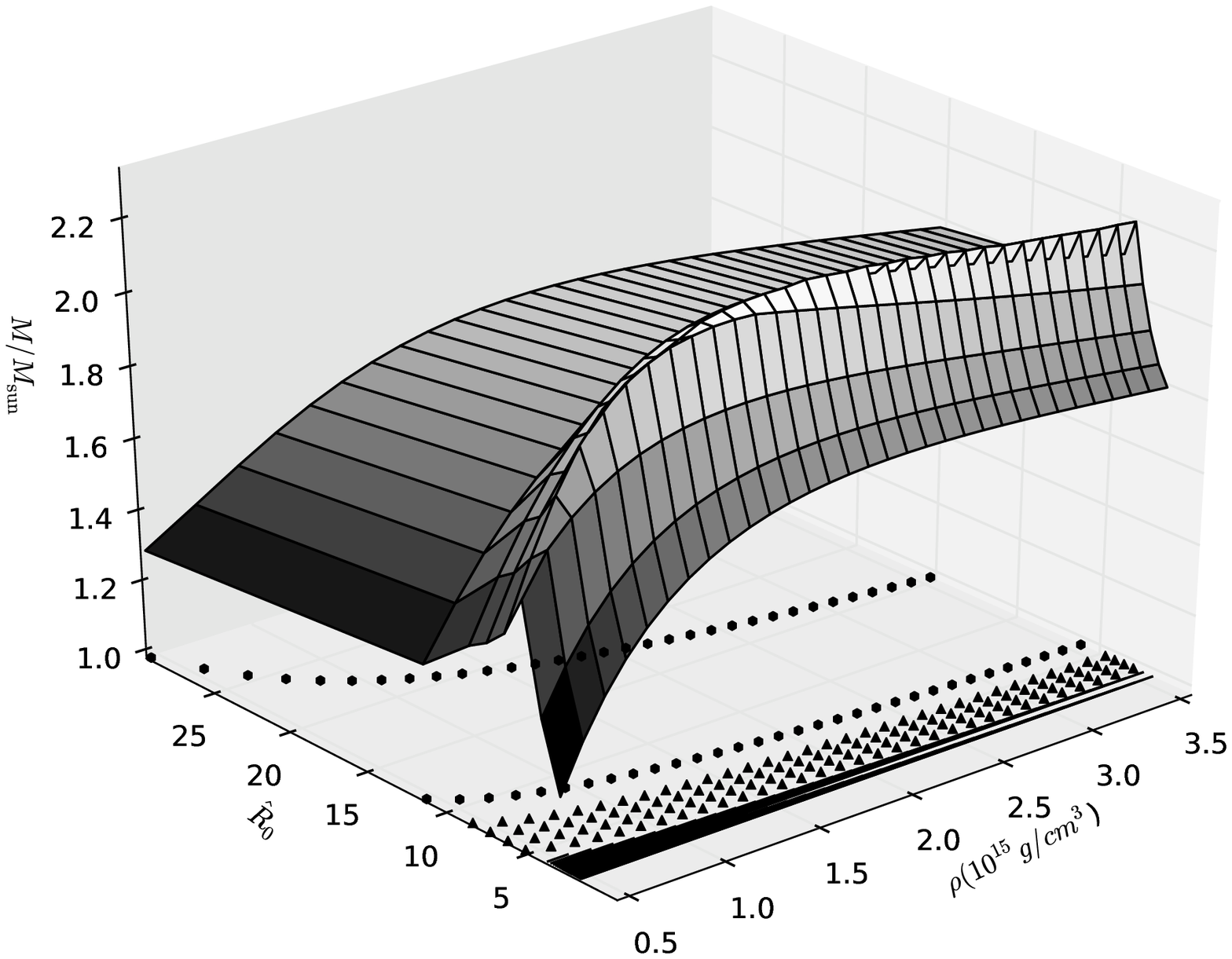}
  \includegraphics[scale=0.45]{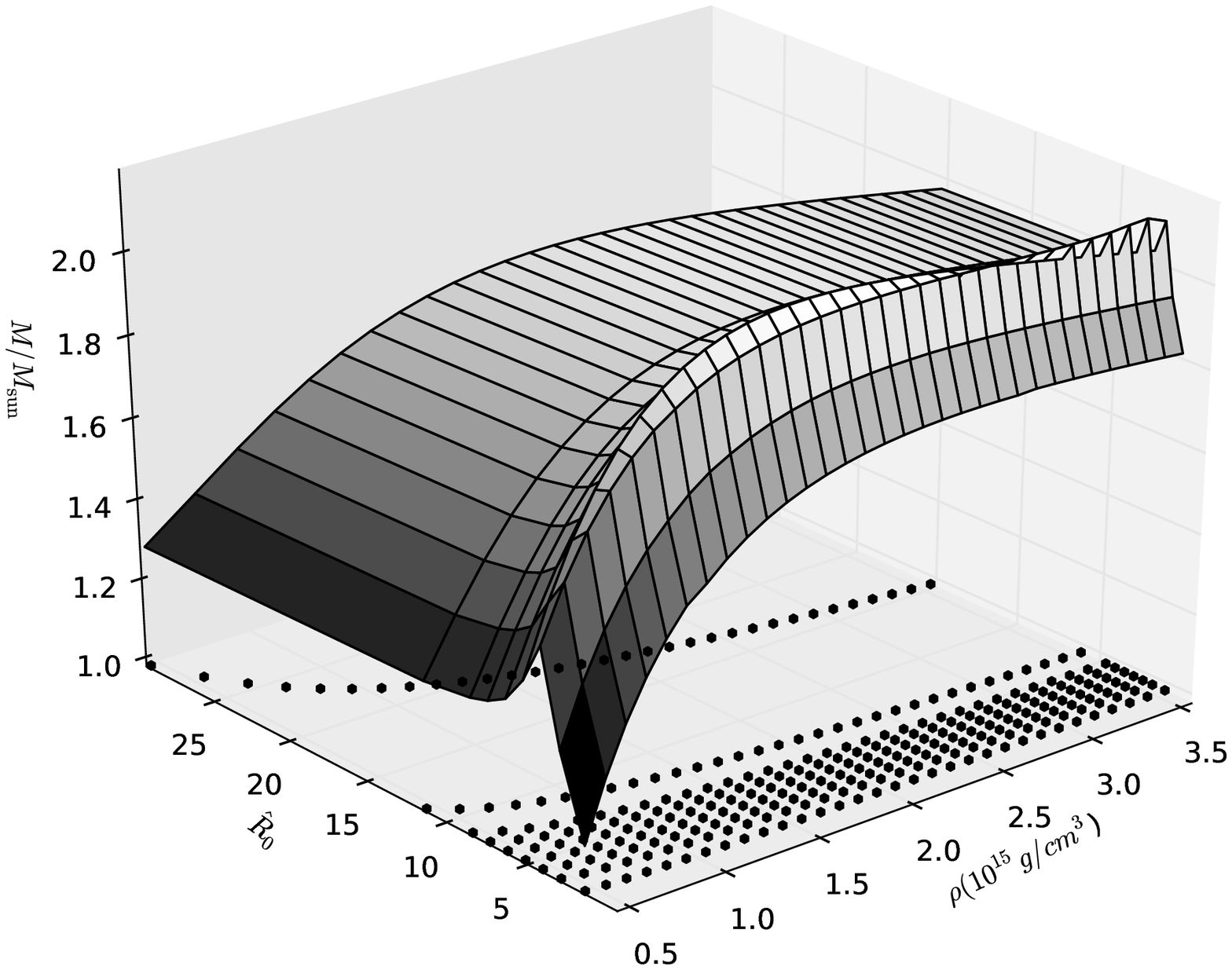}  
  \end{center}
  \caption{Two-dimensional surfaces of equilibrium models
 of differentially rotating neutron stars
 embedded in the space
 $\hat R_0-\rho_{max}-M/M_{sun}$. 
 The first panel is for  $\alpha=-1$, the second one is for
 $\alpha=-2$ and the last one is for $\alpha=-4$.
 The lines at the bottom of each panel are drawn to clarify
 which of the three terminal conditions (see Section~\ref{subsec: numerical sequence}) 
 of sequence applies. Triangles are for
 sequences reaching $T/|W|=0.14$, dots are
 for sequences terminating with mass
  shedding and thick solid lines are for the case when topology change
 occurs.}
  \label{fig:Mass surface plot}
\end{figure}

To visualize our results, we plot gravitational mass of stellar models in the
parameter space. Figure \ref{fig:Mass surface plot} shows the
surfaces of equilibrium models embedded in the 
$R_0-\rho_{max}-M/M_{sun}$ space, where $M$ is the gravitational mass in units of solar masses, for
three different values of the parameter $\alpha$: $\alpha=-1$ (top panel),
$\alpha=-2$ (middle panel) and $\alpha=-4$ (bottom panel).  
In order to have the same degree of differential rotation for all the models with a 
constant $R_0$, following the prescription of \citet{Shapiro2000}, we define:
\begin{equation}
\hat R_0=R_0/R_c,
\end{equation}
where $R_c$ is the circumferential radius.
 
On the bases of these
plots, we marked each parameter pair $(\rho_{\rm max}, \hat R_0)$ with
different symbols to indicate how the equilibrium sequence ends.
The triangles correspond to the parameters for which 
$T/|W|$ of the model reaches $0.14$ (and the computations are stopped
there). The dots correspond to the parameter pairs for which the sequence
terminates at the mass-shedding limit. The solid lines are for the parameters for which
the sequence of spheroids ends with a topological change.

As mentioned in Section \ref{subsec: numerical sequence}, our sequences terminate
at the three different criteria. Therefore one needs to be careful in
reading these figures. It is important to stress that for the sequences terminating 
at $T/|W|=0.14$ and at the topological change, the value of mass plotted here
is \emph{not} the maximum mass of the sequence. The solutions of the toroidal class
do not possess a value for the maximum mass, instead this quantity can arbitrarily increase 
as the torus becomes thinner and thinner along the sequence of equilibrium figures (see. \citet{Ansorg2003}).

It is seen from Eq.(\ref{eq: functional form of g}) that 
the angular velocity profile becomes uniform as $\hat R_0\to\infty$.  In this rigid rotation limit,
the sequence of $N=1$ polytropes is known to terminate with mass-shedding state before
reaching $T/|W|=0.14$. The degree  of differential rotation depends both on the power-law exponent $\alpha$ and on the radius $R_0$. We may regard it as \emph{weak} when the profile is close to uniform rotation (i.e. $R_0 \to \infty$ and/or $\alpha \to -\infty$).  As in the case of uniform rotation, for stars with weak differential rotation the centrifugal force at the equator on the surface reaches a value at which the star sheds mass before its $T/|W|$ value reaches $0.14$. In this case all the sequences terminate at mass-shedding limit before reaching $T/|W|=0.14$ or the point of topology change. However a star with sufficiently 
strong differential rotation can store large rotational energy deep inside the star
and allow the surface angular frequency to be smaller than that of the mass-shedding limit.
Therefore a star with $\alpha=-1$ or $-2$ reaches the critical point 
$T/|W|=0.14$ before encountering mass-shedding or topology change, provided that $\hat R_0$
is small enough (that is a smaller core region of nearly uniform rotation). 
For these choices of $\alpha$ and sufficiently small $\hat R_0$, 
we also see the appearance of topological change {\it before} the critical 
$T/|W|=0.14$ is reached (represented as the triangles in 
Fig.\ref{fig:Mass surface plot}). Indeed, the sequences which we terminate
at $T/|W|=0.14$ will eventually see the topology change if extended to faster 
rotation.

An important observation here is that for $\alpha=-4$ none of the models 
reach $T/|W|=0.14$ before mass shedding occurs, and
for $\alpha=-1,$ and $-2$ we need to set $\hat R_0$ sufficiently small (i.e., enabling high
degree of differential rotation) to have the critical $T/|W|$ before
topology change or mass shedding occurs.

\section{Summary and Discussion}
We have introduced a new rotation profile to study 
equilibrium sequences of differentially
rotating relativistic stars. Compared with the previous studies,
which assume only one type of rotation profile,
we are now able to investigate a broader class of rotating stars.

As a first step towards systematic studies, we focus on the simplest
neutron star model with a polytropic EOS with index $N=1$.
For the rotation profiles that allow analytic expressions
in the Eq. (\ref{hydrostatic eq}), we computed
sequences of spheroidal equilibria that start from non-rotating
stars. Special attention is paid to the appearance of the critical
point to the secular instability (CFS or viscous instability) to
bar-shaped deformation of the star. We see that the appearance
of the critical point may occur for rather strong differential rotation
in the case of $\alpha=-1$ and $-2$ in Eq.(\ref{eq: functional form of g}),
which corresponds respectively to so-called ``j-const." and ``v-const." rotation
profiles in Newtonian stars \citep{EriguchiMueller1985}.
Even in these cases, the critical point does not appear before
the configuration changes from a spheroidal topology to a toroidal one,
unless the parameter $R_0$ in Eq.(\ref{eq: functional form of g}) 
is sufficiently small.

When $\alpha=-4$ instead there seems no critical point of CFS instability before 
the equilibrium sequences reach their mass-shedding limit. This last case is relevant 
since it appears to mimic the approximate rotation profile of some of the quasi-stationary
HMNSs seen in the numerical simulations of neutron star mergers.
However these stars may be susceptible to the relatively new class
of dynamical instability (so-called "low $T/|W|$ instability", e.g. \citet{Centrella_etal2001, Shibata_etal2002, Watts_etal2003, SaijoYoshida2006, OuTohline2006, Baiotti2008}),
whose existence relies on the strong shear flow (\citet{Watts2005,Corvino2010}). 
Indeed, numerical simulations of binary NSs do indicate that the HMNS develops a bar deformation (e.g.\citet{ShibataUryu2002}; \citet{Baiotti2008}) on top of the background star with approximate axisymmetry. It would be interesting to study the appearance of low $T/|W|$ instability
by using our equilibrium stars with different rotation profiles since the survival time of the merger 
remnant could be an important diagnostic. The observation of the delay between detection of the gravitational wave signal form a binary neutron stars merger and the possible observation of the short gamma ray burst counterpart would then help to put constraints on the internal structure of a NS.
		     		     
\begin{acknowledgements}
We thank Marcus Ansorg and Carlos Palenzuela for useful discussions. 
This work is supported in part by a Grant-in-Aid for Scientific 
Research (C) of Japan Society for the 
Promotion of Science (20540225).
FG wants to thank Aaryn Tonita for the essential 
support in this work, Sam Lander for the fruitful 
discussions and finally Luciano Rezzolla for 
the continuous support that he devoted in this project.
\end{acknowledgements}


\appendix
\section{Integration of $g(\Omega)$}
For some interesting cases, we have an analytic expression for the 
integral of $g(\Omega)$ which appears in Eq.(\ref{1st integral of hydrostatic eq}).

We define a parameter $k:=(R_0\Omega_c)^2$ and normalize $\Omega$
as $\xi:=\Omega/\Omega_c$. Then the integration of the functional $g(\Omega)$ is written as
\begin{equation}
	\int g(\Omega)d\Omega 
	= \int \frac{k\xi(\xi^\alpha-1)}{1-k\xi^2(\xi^\alpha-1)}d\xi.
\end{equation}

For $\alpha=-1$, the right hand side reduces to
\begin{equation}
	\sqrt{\frac{k}{4-k}} {\rm Arctan}\left(
	\sqrt{\frac{k}{4-k}}(2\xi-1)\right)
	-\frac{1}{2}\ln(1+k\xi(\xi-1)).
\end{equation}
For $\alpha=-2$, it reduces to
\begin{equation}
	\frac{\ln(1+k(\xi^2-1))-2k\ln\xi}{2(k-1)}.
\end{equation}
Finally, for $\alpha=-4$ we have
\begin{equation}
	-\ln\xi + \frac{1}{2\sqrt{1+4k^2}}\cdot
	\ln\left(\frac{-1+\sqrt{1+4k^2}-2k\xi^2}{+1+\sqrt{1+4k^2}+2k\xi^2}
		\right).
\end{equation}

\bibliographystyle{aa} 
\bibliography{myBib} 

\end{document}